# Investigating Population Dynamics of the Kumbh Mela through the Lens of Cell Phone Data

Jukka-Pekka Onnela[1] and Tarun Khanna[2]
[1] Department of Biostatistics, Harvard T.H. Chan School of Public Health,
655 Huntington Avenue, Boston, MA 02115
[2] Harvard Business School, Soldiers Field Road, Boston, MA 02163;
Harvard South Asia Institute, 1730 Cambridge Street, Cambridge, MA 02138

**Abstract**
The Kumbh is a religious Hindu festival that has been celebrated for centuries. The 2013 Kumbh Mela, a grander form of the annual Kumbh, was purportedly the largest gathering of people in human history. Many of the participants carried cell phones, making it possible for us to use a data-driven approach to document this magnificent festival. We used Call Detail Records (CDRs) from participants attending the event, a total of 390 million records, to investigate its population dynamics. We report here on some of our preliminary findings.

**Introduction**
The Kumbh Mela is a religious Hindu festival that has been celebrated for hundreds of years. On the broad sandy flats left behind by the receding waters of the rainy season, a temporary "pop-up" megacity is constructed in a matter of weeks to house the festival. The Kumbh Mela has evoked endless fascination and scrutiny by observers dating back to Chinese and Arab itinerant scholars from centuries past, to contemporary academics, mostly social scientists and humanists of various disciplinary persuasions. Both participants and external observers have documented past Kumbh festivals, and this book brings together various perspectives using different methodologies to the study of the 2013 Kumbh Mela.

In this chapter, we examine the use of cell phone communication metadata to study the 2013 Kumbh Mela. Telecom operators worldwide collect these data routinely for billing and research purposes. At bare minimum, metadata contain information on who calls whom, at what time, and for how long. The metadata studied here, also known as Call Detail Records (CDRs), were made available to us by the Indian telecom operator Bharti Airtel for the period from January 1, 2013 to March 31, 2013. The cohort we introduce here consists of all Bharti Airtel customers who were present at the Kumbh Mela site during this time period and used their cell phone at least once to communicate with someone, whether as senders or receivers of communications. We describe the details of this cohort, consisting of approximately 390 million communication events. To put this number into perspective, if a single individual were to take just one second to examine each record, it would still take twelve years to go through the data (without breaks).

Cell phone metadata in general, and these data in particular, have a longitudinal, spatial, and network component to them, and they describe both calls and text messages. These data characteristics place our work firmly in the "big-data" context and enable us to address questions that would be difficult or impossible to tackle without these data and the requisite data analytic techniques. We hope that our approach could bring the study of this spectacular historic event—purportedly the largest gathering of people in human history—on the intellectual radar of a host of quantitative scientists, such as physicists and statisticians, who may not typically study these types of cultural phenomena. We should note that given their nature, these data could be analyzed using a broad range of sophisticated statistical techniques and mathematical models. In this chapter, our goal is to introduce the idea of using these data to study the 2013 Kumbh Mela, to put forward what may be attained by this approach, to describe the festival's metadata cohort and some salient features of the data, and to report on some of our preliminary findings. We leave more detailed and sophisticated analyses of these data for future work.

**Data-Driven Study of Human Behavior**
Research on human social interactions has traditionally relied on observations reported by humans, and both self-reported data and observer-recorded data, with varying degrees of observer involvement being used to quantify social interactions[1]. Analyzing human behavior based on electronically generated data has recently become popular. One approach is based on actively "instrumenting" study subjects by having them wear electronic devices, such as sociometers (credit card sized devices worn around the neck), to collect detailed data on various dimensions of social interaction, such as talkativeness and physical proximity to others.[2–4]

An alternative approach is to rely on passively generated electronic records of behavior, such as CDRs that are collected by operators for billing and internal research purposes. These data can be analyzed to investigate the structure of social networks,[5,6] human mobility patterns,[7,8] and the role of geography in constraining social ties and social groups.[9,10] They have also been applied to study infectious diseases in epidemiology.[11] Although active instrumentation, such as the use of sociometers, can result in richer behavioral data, it does not scale well to the study of groups larger than perhaps a few hundred subjects. The passive approach relying on CDRs, in contrast, can in principle scale up to any level and has been used to construct, analyze, and model society-wide social networks.[5,12,13]

The Kumbh Mela has always fascinated scholars across different fields, and often these scholars have focused on studying specific aspects of the gathering, such as the nature of gift giving from one participant to the other or the style of clothing worn by participants. These types of studies are fascinating in their own right, but they necessarily involve many subjective decisions to be made by the investigators on issues like which gift-giving event to document or which dress to photograph. This necessarily requires some preconception of what types of behaviors should be captured, naturally leading to some pre-screening on which behaviors are recorded.

Much of the recent scholarship on the Kumbh Mela (e.g., *Pilgrimage and Power*[14] and the *Kumbh Mela Pop-up Megacity Business Case Study*[15]) has the central construct of the value and nature of information flows at the festival. Information was vital through history for several disparate reasons. During colonial and pre-colonial times, the Mughals and the British Raj saw Allahabad (or Prayag as it was known at one point) as a logistically important town, and the rush of pilgrims often compromised their ability to control information (and to otherwise mobilize the military). Also during the Raj era, the British were concerned that seditious messages might inadvertently be spread far and wide across India after visitors met at the Kumbh Mela and exchanged ideas with content that was not possible to police. In independent India, in contrast, the Kumbh often features as a way for India's leaders to broadcast nationalist messages, as it does in modern India for those politicians who wish to identify with the predominantly Hindu component of the electorate. Whether monitored in days of lore by *sadhus* (wandering holy men), or more currently by modern transient merchants, or by people in neighboring cities, information flows are vital in controlling commerce. Finally, of course, information is key to the prospect of managing the religious event in real time. Indeed, our use of cell phone metadata to shed light on information flows is perhaps most useful for such management of the event in the years to come. We offer our approach in that preliminary spirit.

Cell phones are now ubiquitous in India, and it is perhaps surprising to some readers how many pilgrims traveled with their cell phones to the Kumbh Mela. This fact enables us to adopt a data-driven approach that makes it possible to investigate not only those phenomena that were deemed interesting a previously, but also to capture unanticipated phenomena. We therefore approach the Kumbh as an organic, collective event with as few preconceptions as possible. It should be stressed that the type of approach we advocate here is not a panacea by any means and it clearly has its limitations. We do, however, think that the proposed approach is useful for giving us the proverbial 30,000-foot perspective to this magnificent cultural event and it may very well help complement some of the many other ways used to study the Kumbh Mela.

**Call Detail Records**

Since the data we use to investigate the Kumbh is generated by cell phones and recorded by operators—to understand the potential and limitations of this approach, it is helpful to understand the basics of what happens when a person uses his or her cell phone to call another user. In its most basic form, a cell phone functions like a two-way radio containing a radio receiver and a radio transmitter that transmits the communication data to the nearest cell tower. The cell tower is a steel pole or lattice structure that rises into the air and contains structures to support antennas. Cell towers are accompanied by base stations that contain the needed electronics for handling data transmission. Base station cell sites themselves are connected to the Mobile Telephone Switching Office (MTSO) that monitors traffic and arranges hand-offs between base stations. Each cell phone and nearest base station keep track of the cell phone that happens to be located at a given time. When a cell phone user dials another user, the MTSO searches its database for the target number and transmits the call to the cell tower that is nearest to the person called. The areas of signal coverage from adjacent cell towers overlap slightly near the border of the cells, and if a phone moves to the edge of a cell, the base stations monitor and coordinate with each other through the MTSO and arrange "hand-offs" from one cell tower to another to ensure that the call does not get disrupted.

A CDR is a digital record gathered from an instance of a telecommunications transaction described above. It is automatically generated by the mobile network operator, and contains specific data points about the transaction, but not the content of the transaction, such as voice call and text message data. The types of information included in a CDR may include the following: the phone number of the subscriber originating the call, the phone number receiving the call, the starting time of the call (date/time), the duration of the call, a unique sequence number identifying the record, the identification of the telephone exchange or equipment writing the record, the route by which the call entered the exchange (the nearest cell phone tower of the initiating party), the route by which the call left the exchange (the nearest cell phone tower of the receiving party), and the event type (voice call or SMS).[16]

Table 1 illustrates what simple CDR data (artificial, in this case) look like. The first three columns specify the date, the start time of the event, and the type of event. The fourth column indicates call duration (for voice calls only). The fifth and sixth columns give anonymized (surrogate) IDs for the caller and recipient, respectively. The last two columns in the table indicate what cell towers were used to transmit and receive the communication event, enabling one to geographically locate individuals at the time of communication. It is this geographical component of the metadata that enables us to define a cohort that is spatially localized at the Kumbh Mela site.

Table 1. An example of the metadata used in the study. *VC* refers to voice call and *SMS* to text message.

| Date | Start time | Event type | Duration (s) | Caller ID | Callee ID | Caller tower ID | Callee tower ID |
|---|---|---|---|---|---|---|---|
| 09/02/2013 | 14:36 | VC | 185 | 9912345678 | 4412345678 | 54 | 42 |
| 10/02/2013 | 14:41 | VC | 252 | 9912345678 | 6612345678 | 42 | 73 |
| 11/02/2013 | 14:42 | SMS | NA | 4412345678 | 6612345678 | 73 | 73 |

**The 2013 Kumbh Cohort**

The 2013 Kumbh cohort consists of any Bharti Airtel customer who used his or her cell phone either to make or receive a call, or to send or receive a text message at the Kumbh Mela site between January 1 and March 31, 2013. The Kumbh site comprises 207 unique cell sites, or cell towers, located within the area. This larger than normal density of towers was accomplished by adding several mobile cell towers, situated in fixed locations throughout the festival to meet the expected high service demand. Because a communication event (call or SMS) occurs at the level of pairs of individuals, it is useful to elaborate what is and what is not observed as a consequence of this study design. Any communication event involves two parties—Person A and Person B. Person A may or may not be an Airtel customer, and may or may not be present at the Kumbh Mela site during the period of investigation; the same holds true for Person B. Therefore, each interaction involves zero, one, or two Airtel customers, and either of the two people may or may not be present at the Kumbh site. We observe any communication event that involves at least one Airtel customer who is present at the Kumbh site during the period of investigation. The other party may be a customer of any operator (Airtel or non-Airtel) and may be physically located anywhere in the world (at the Kumbh site or outside of it). Figure 1 shows a schematic of the study design.

We note that one can divide all communication ties into intra-Kumbh communications (both people at the venue) and extra-Kumbh communications (only one person at the venue). We speculate that the intra-Kumbh ties are likely used for local coordination, which will become more important as population density increases and crowds rise, whereas the extra-Kumbh ties are likely driven by other factors. Future research should investigate the differences between these two distinct types of communication events and their consequences for population dynamics.

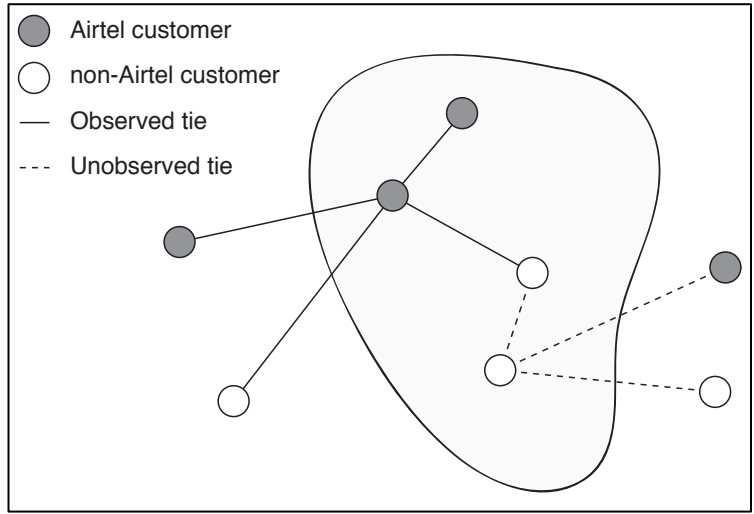

**Figure 1. Schematic of the study design. The large shaded area in the middle represents the Kumbh Mela site. Airtel customers are represented with dark nodes and non-Airtel customers with white nodes. A tie connecting a pair of nodes corresponds to a communication event (call or SMS); observed ties are shown as solid lines and unobserved ties as dashed lines. We observe any communication event that involves at least one Airtel customer who is at the Kumbh site during the investigated time period.**

The following data were available for each communication event for the individuals in the Kumbh Mela cohort:
- Number of the caller
- IMSI number
- CDR type (call or SMS)
- Number of the callee
- Cell tower ID
- Date
- Time
- Call duration

The IMSI number corresponds to what is often called the phone's SIM card number; Cell tower IDs are codes used to associate particular cell towers with each communication event.

Privacy of human subjects is always an important concern in any study. In our past work with cell phone metadata[5,6,10] we have relied on anonymized data, which involves mapping the phone numbers and other identifiers to surrogate keys in such a way that this mapping cannot be reversed. The goal of our past and ongoing research efforts is to obtain generalizable knowledge about human behavior, thus an individual identifier is a nuisance parameter of no scientific interest and yet, importantly, it poses an unnecessary threat to subjects because of the potential risk of subject re-identification. Consequently, data anonymization upholds the scientific value of such data, while simultaneously providing excellent protection for human subjects. Anonymization is now a routine procedure in this line of work and something that is mandated by most university Institutional Review Board (IRB) regulations.

In the present project, the authors did not have direct access to the data. Instead, the data resided at all times on Airtel's computers in Gurgaon, India. The research was carried out by sending computer-programming scripts to an Airtel employee who ran the analyses on a dedicated computer and sent the results, consisting of summary statistics of interest back to the authors.

**Preliminary Findings**

In the following section, we report on some preliminary findings pertaining to intensity of communication activity, daily fluctuations in cohort size, cumulative cohort size, and length of stay.

**Intensity of communication**

As a starting point to our analyses, we investigated the total volume of communication in the cohort. The 2013 Kumbh cohort exchanged a total of 146 million (145,736,764) text messages and 245 million (245,252,102) calls, resulting in a total of 390 million (390,988,866) communication events. When plotted as a function of time in Figure 2, where the $x$-axis gives the index of the day with January 1, 2013 corresponding to day 0, it is clear that the number of calls and text messages are not uniformly distributed in time but instead display one prominent and several smaller peaks.

The main bathing day of the 2013 Kumbh Mela was February 10, 2013, corresponding to day index 40, and this is precisely the location of the most prominent voice call and text messaging peaks. In addition, calling activity was above its baseline value for about a week leading to the main bathing day and then declined over the next several days. The number of text messages also had a clear peak surrounding this event, but it appeared more concentrated around the main bathing day, rising closer to the event than the curve for phone calls and declining faster after it. We report the daily text message and voice call counts surrounding the main bathing day in Table 2.

The ramping up to the peaks on significant bathing days can provide useful information for capacity planning of all sorts, as well as for monitoring (human and vehicular) traffic congestion patterns. These types of regularties in the data could be used for future predictive modeling purposes.

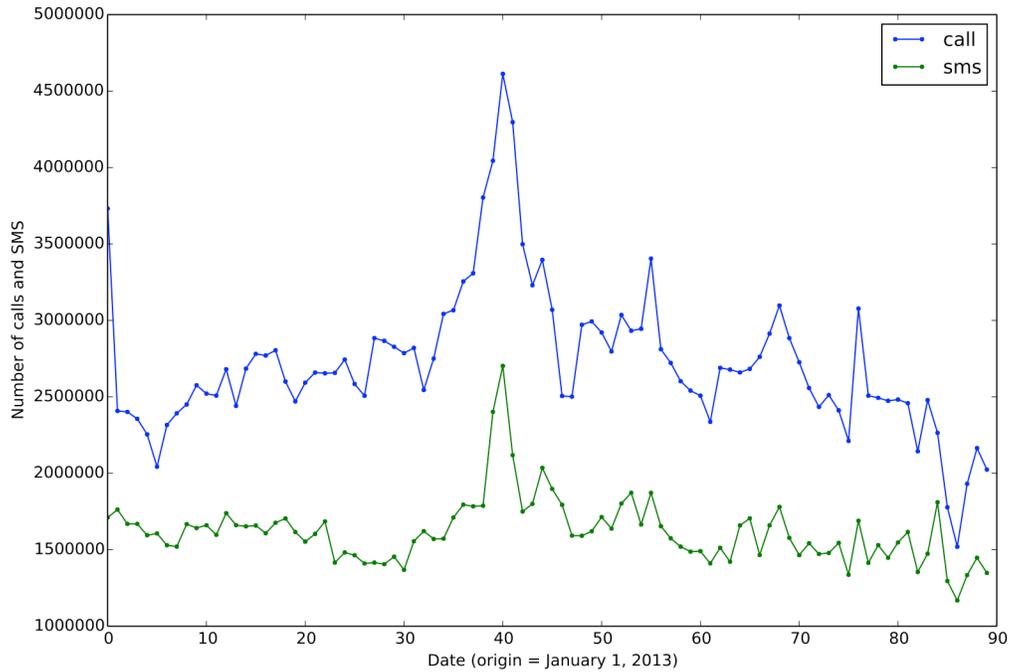

**Figure 2. The number of calls and text messages as a function of time in the 2013 Kumbh cohort.**

| Date | SMS count | Call count | Total count |
|---|---|---|---|
| 7-Feb-13 | 1,784,464 | 3,308,496 | 5,092,960 |
| 8-Feb-13 | 1,788,094 | 3,803,392 | 5,591,486 |
| 9-Feb-13 | 2,401,666 | 4,044,101 | 6,445,767 |
| **10-Feb-13** | **2,702,578** | **4,612,422** | **7,315,000** |
| 11-Feb-13 | 2,118,802 | 4,296,436 | 6,415,238 |
| 12-Feb-13 | 1,750,374 | 3,498,322 | 5,248,696 |
| 13-Feb-13 | 1,800,575 | 3,230,284 | 5,030,859 |

**Table 2. SMS, call, and total event count for the days surrounding the main bathing day of February 10, 2013. This day commands the highest communication volume.**

## Daily fluctuations in cohort size

Cell phone metadata can be used to investigate the number of people in attendance at the 2013 Kumbh. Here it is useful to distinguish between daily fluctuations in the size of the Kumbh cohort and its aggregate or cumulative size from January 1, 2013 to March 31, 2013. We start with the daily fluctuations by plotting the number of unique daily SIM cards in Figure 3. In this plot, each point represents the number of unique SIM cards that were used to communicate that day, which are re-aggregated each day of the investigation period. If a SIM card is not used on a given day, it does not contribute towards that day's count, and its use on other days is irrelevant for that day's count. One can immediately observe the major bathing days in this plot, with the February 10 bathing day (at time index 40) clearly standing out from the other peaks. On that day, approximately 800,000 unique SIM cards were active.

Using the above numbers for determining the number of Kumbh visitors depends, among other factors; on the fraction of people carrying cell phones and to what extent that fraction may change in time. While estimating the precise numbers is part of our ongoing work, we want to point out that as long as the fraction of people with cell phones and the frequency of cell phone use remain more or less constant, one can use these findings to study day-to-day changes in the size of the Kumbh cohort. For example, from the perspective of infectious disease management, it is important to have accurate estimates of the size of the population and, in particular, of the changes in its size. This is for the simple reason that if the number of cases of a given disease goes up by, say, 50% from one day to the next, if there is an accompanying change of 50% in the size of the underlying population, we likely do not have an epidemic. In contrast, if the population size remains constant, we may have an epidemic at hand. For epidemiological applications, such as detecting an impending epidemic, learning about *relative changes* in cohort size is often sufficient. The present approach may, in fact, be more informative to that end than estimates based on more conventional sources, such as traffic monitoring to and from the site.

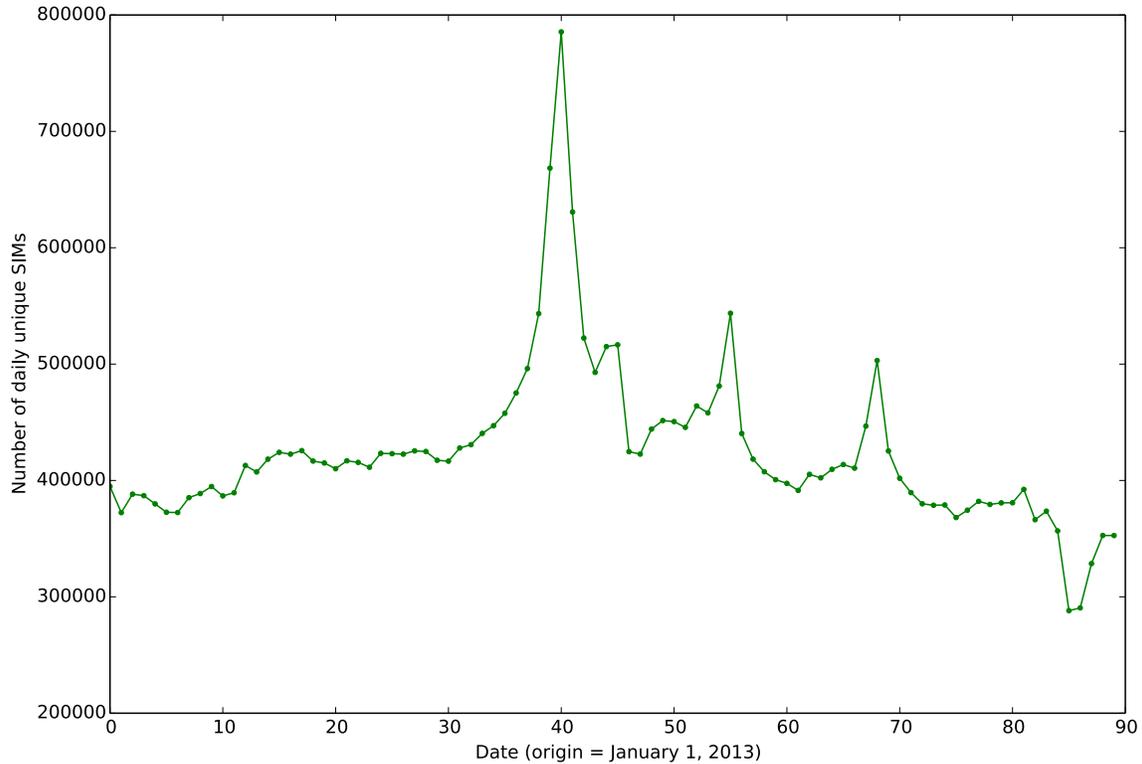

**Figure 3. The number of unique SIM cards in use each day.**

**Cumulative cohort size**

We focused above on daily changes in the number of individuals (SIM cards) present at the Kumbh Mela site and argued that cell phone metadata and their statistical analysis can yield important insights into this problem. There are a number of other approaches that can be used to study the number of people present at any given location, such as aerial observation or satellite imaging of the venue. The challenge with these types of approaches is that if the gathering lasts for several weeks like the Kumbh does, it becomes difficult to estimate the aggregate number of people who have passed through the event. For example, we reported above that approximately 800,000 unique SIM cards were active on the main bathing day. However, there is no way to tell from plots like Figure 3 whether we have the same set of visitors across days—or whether there is an inflow and a corresponding outflow of people that keeps the number of daily visitors more or less constant but actually corresponds to an entirely different population dynamic that contributes to an increasing cumulative number of visitors.

To distinguish between the number of daily visitors and the cumulative number of visitors, we examine the cumulative number of unique SIM cards over the time period from January 1st to March 31st. In Figure 4, we plot the number of unique SIM cards that have been active at least once by the time index given on the x-axis. By March 31st, approximately 3.9 million SIM cards had been used by the cohort. In a cumulative plot such as this one, no two adjacent points are independent, which on the one hand leads to a smooth curve, but on the other hand makes it more difficult to detect changes in underlying population dynamics. That said, the three major bathing days are visible as "shoulders" in both the curves, suggesting that large groups of people arrived to the venue specifically to experience these major bathing days.

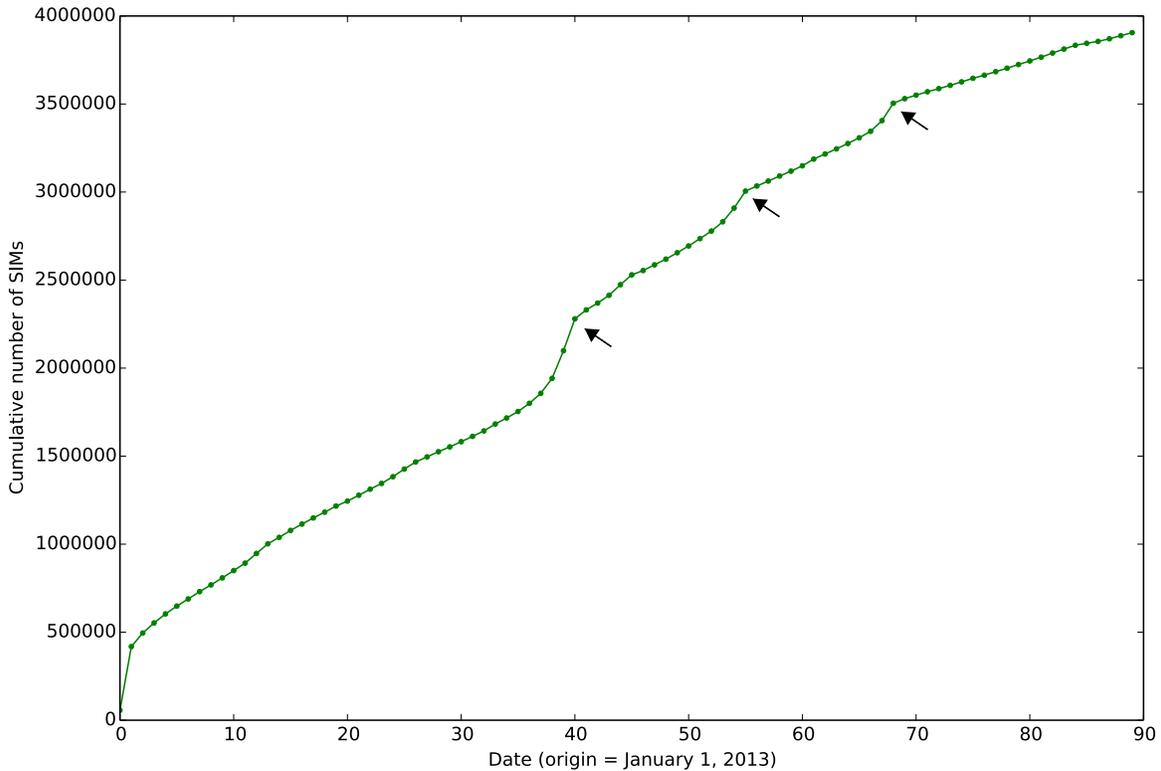

Figure 4. The cumulative number of unique SIM cards in the 2013 Kumbh cohort up to the point in time given on the x-axis. The arrows point to "shoulders" in the curve, corresponding to major bathing days.

**Length of stay**

There are many different categories of visitors to the Kumbh Mela, ranging from the curious visitor in passing, to the deeply religious *ascetic* staying for the entire duration of the event. To investigate the length of stay of the visitors is therefore an interesting question in its own right, but is also helpful for studying the population churn or for estimating the total number of individual visitors throughout the entire event. Here we focus on the length of stay question by recording for each SIM card the time $t_a$ when it was first used at the venue and the time $t_b$ when it was last used. We define length of stay as $d = t_b - t_a$.

In examining this number, we need to consider the fact that any given SIM card may have been present at the venue before $t_a$ but was not used for the first time until at time $t_a$. Similarly, a SIM card could have remained at the venue after $t_b$ without being used subsequently. For this reason, the length of stay is estimated as a simple difference between $t_b$ and $t_a$ resulting in an interval-censored estimate. It can be interpreted as a lower bound for the actual (unobserved) duration of stay, leading to conservative estimates. Using this approach, we found the mean length of stay to be 18.1 days, the minimum and maximum being one day and eighty-nine days, respectively. There was also substantial variability in the lengths of stay, with a standard deviation of 29.6 days. Overall, 23.8% of the stays were longer than thirty days and 15.0% of the stays were over sixty days.

We note that while it is possible that the same SIM card is used in multiple handsets, but as long as the SIM card arrives and leaves the site when the person does, there is no systematic bias introduced to our length-of-stay estimates. There are however other scenarios where a bias could be introduced. For example—a person could leave the Kumbh Mela site but give their SIM card to a friend who is staying longer, resulting in an overestimated of length of stay. Considering joint SIM card numbers and unique handset numbers could alleviate this and other potential concerns, more sophisticated statistical approaches could be employed to incorporate SIM card sharing and other behavioral scenarios into the estimation process, leading to more accurate and robust estimates.

**Future Research**

In this section, we have reported on a novel approach that may be used to study population dynamics of individuals at the Kumbh Mela and have reported on some of our preliminary results. We would like to conclude this section with two messages. First, from the early results presented here, it is clear that this overall approach can be used in a number of contexts where large numbers of people have gathered together. While the Kumbh Mela is obviously a very carefully planned event, the approach is also suitable to study the population dynamics of misplaced populations such as those resulting from natural disasters like earthquakes or floods. Second, it is clear that the richness of the cell phone metadata lends itself to very sophisticated statistical and mathematical analyses that can incorporate dimensions of the data we have not utilized here, such as the properties of the social network constructed from cell phone call patterns, or the spatial dynamics of individuals at the Kumbh Mela venue obtained from cell tower data. These types of analyses could shed light on large-scale crowd behavior and would be helpful for understanding and predicting crowd movements, thereby helping with crowd control and traffic planning; measures that could help avoid congestion and human stampedes. We leave the investigation of these other topics for future work.

**Bibliography**

1. Onnela, J.-P., Waber, B., Pentland, Alex (Sandy), Schnorf, S. & Lazer, D. Using sociometers to quantify social interaction patterns. *Sci. Rep.* **4,** (2014).

2. Olguin Olguin, D. *et al.* Sensible organizations: technology and methodology for automatically measuring organizational behavior. *IEEE Trans. Syst. Man. Cybern. B. Cybern.* **39,** 43–55 (2009).

3. Eagle, N. & (Sandy) Pentland, A. Reality mining: sensing complex social systems. *Pers. Ubiquitous Comput.* **10,** 255–268 (2005).

4. Cattuto, C. *et al.* Dynamics of person-to-person interactions from distributed RFID sensor networks. *PLoS One* **5,** e11596 (2010).

5. Onnela, J.-P. *et al.* Structure and tie strengths in mobile communication networks. *Proc. Natl. Acad. Sci. U. S. A.* **104,** 7332–6 (2007).

6. Onnela, J.-P. *et al.* Analysis of a large-scale weighted network of one-to-one human communication. *New J. Phys.* **9,** 179–179 (2007).

7. González, M. C., Hidalgo, C. A. & Barabási, A.-L. Understanding individual human mobility patterns. *Nature* **453,** 779–82 (2008).

8. Calabrese, F., Diao, M., Di Lorenzo, G., Ferreira, J. & Ratti, C. Understanding individual mobility patterns from urban sensing data: A mobile phone trace example. *Transp. Res. Part C Emerg. Technol.* **26,** 301–313 (2013).

9. Lambiotte, R. *et al.* Geographical dispersal of mobile communication networks. *Phys. A Stat. Mech. its Appl.* **387,** 5317–5325 (2008).

10. Onnela, J.-P., Arbesman, S., González, M. C., Barabási, A.-L. & Christakis, N. A. Geographic constraints on social network groups. *PLoS One* **6,** e16939 (2011).

11. Wesolowski, A. *et al.* Quantifying the impact of human mobility on malaria. *Science* **338,** 267–70 (2012).

12. Blondel, V. D., Guillaume, J.-L., Lambiotte, R. & Lefebvre, E. Fast unfolding of communities in large networks. *J. Stat. Mech. Theory Exp.* **2008,** P10008 (2008).

13. Ratti, C. *et al.* Redrawing the map of Great Britain from a network of human interactions. *PLoS One* **5,** e14248 (2010).

14. Maclean, K. *Pilgrimage and Power: The Kumbh Mela in Allahabad, 1765-1954*. (Oxford University Press, 2008).

15. Khanna, T., Macomber, J. & Chaturvedi, S. Kumbh Mela, India's pop-up Megacity. *Harvard Bus. Publ. Case Study* (2013).

16. Horak, R. *Telecommunications and Data Communications Handbook [Hardcover]*. 832 (Wiley-Interscience; 2 edition, 2008).

**Acknowledgements**

The authors wish to acknowledge the help of Jeanette Lorme. Onnela is supported by a 2014 Career Incubator Award by the Harvard T.H. Chan School of Public Health; Khanna is supported by the Harvard Business School.